\documentclass[12pt,a4paper]{article}
\usepackage[margin=2.5cm]{geometry}

\usepackage{amsmath}
\usepackage{amssymb}
\usepackage{xcolor}
\usepackage{graphicx}
\usepackage{booktabs}
\usepackage{url}
\usepackage{hyperref}

\begin{document}

\title{Fast electrons and high energy photons produced in strong-field ionization of atoms as diagnostics of multi-petawatt laser beams}

\author{%
\parbox{\textwidth}{\centering A.V. Khudov$^{1}$, N.V. Makarenko$^{2}$, A.A. Mironov$^{3}$, S.V. Popruzhenko$^{2}$}\\[8pt]
\parbox{\textwidth}{\centering\small $^{1}$ Prokhorov General Physics Institute of the Russian Academy of Sciences, Vavilova 38, 119991, Moscow, Russia}\\[2pt]
\parbox{\textwidth}{\centering\small $^{2}$ National Research Nuclear University MEPhI, Kashirskoe shosse 31, 115409, Moscow, Russia}\\[2pt]
\parbox{\textwidth}{\centering\small $^{3}$ Center for Theoretical Physics (CPHT), CNRS, {\'E}cole Polytechnique, Institut Polytechnique de Paris, 91128 Palaiseau, France}}
\date{}

\maketitle

\begin{abstract}
We calculate numerically energy spectra and angular distributions of photoelectrons and X-ray photons emitted in the process of tunnel photoionization of argon by a focused laser beam of multi-petawatt power.
By scanning the beam power and varying the degree of focusing and the field polarization, we show these distributions to be sensitive both to the peak laser intensity and to the field configuration in the focal spot. 
Investigated parameters of the laser beam correspond to those of several recently commissioned lasers of multi-petawatt power.
These findings, in line with the earlier reported proposals, provide a basis for the development of a reliable ionization diagnostics of extremely intense laser fields.
\end{abstract}

%
%
%
%
%

\section{Introduction}

The development of super-intense laser sources of electromagnetic radiation has demonstrated considerable progress over the past two decades. 
Several multi-petawatt laser facilities have been put into operation \cite{corels,ELI-BL,ELI-NP1,SULF1,apollon_25}.
Even more ambitious projects of sub-exawatt sources are currently under development and construction \cite{XCELS,sel_21}.
Among the multiple potential applications of petawatt (PW), multi-PW and exawatt sources of ultrashort laser pulses, studies of fundamental effects of the interaction of electromagnetic field with charged particles in nonlinear and non-perturbative regimes are of primary significance for fundamental physics.
These include investigations of the radiation reaction force in classical and quantum regimes, relativistic ionization, effects of nonlinear quantum electrodynamics, various schemes of laser acceleration of electrons and ions, and more.
The current status of this thriving research area is outlined in reviews \cite{blackburn_rmpp20,gonoskov_rmp22,fedotov_pr23,popruzhenko_usp23} and the literature quoted there.
The precise knowledge of the peak intensity, which can be achieved in the focus of a super-intense laser beam, and of the intensity distribution and field configuration in the focal area is of key significance for the reliable analysis of experimental data as well as for the determination of principal requirements for the search of certain effects with the use of a given laser source. 

This raises a high demand for methods of direct diagnostics of extreme electromagnetic fields in the foci of multi-PW laser beams.
To clarify terminology, note that we call diagnostics {\em direct} if it is based on measurements of effects, which take place in the laser focus in the full power mode of the beam.
Several nonlinear effects, which can serve as sensitive tools for field diagnostics, have been examined in theory and, partially, experimentally.
An incomplete list includes tunnel ionization of multielectron atoms \cite{walker_pra01,ueda_pra03,link_instr06,poprz_pra19,peganov_mre20,mironov_pre25}, acceleration of electrons and ions created in the laser focus through ionization \cite{vais_apb19,dichiara_prl08,yandow_pra19,yandow_pra24}, nonlinear Thomson and Compton scattering of photons of an intense laser pulse on a beam of fast electrons \cite{dipiazza_ol12,he_oe19,mackenroth_njp19}.
For further references, we direct the reader to review \cite{popruzhenko_usp23}.
Among these effects, tunnel ionization of heavy atoms and ions possesses two significant advantages: it is highly nonlinear and local.
Each of these features potentially allows us to probe the laser intensity and measure its peak value with good accuracy \cite{poprz_pra19,peganov_mre20}.
However, the method has also certain shortcomings: (a) the tunnel ionization diagnostics appears most sensitive to the value of the laser intensity  only when the latter is close to the ionization threshold of s-electrons \cite{mironov_pre25} and (b) in the non-relativistic domain, which for tunnel ionization spreads up to $\simeq 10^{24}-10^{25}{\rm W/cm}^2$ the ionization probability is solely determined by the electric field amplitude so that the method probes intensity and not the electric and magnetic fields vectors.
At the same time, in an experiment targeted at measurements of production of highly charged ions \cite{walker_pra01,ueda_pra03,link_instr06}, one could register also fast electrons produced through nonlinear ionization \cite{yandow_pra19,yandow_pra24} and high-energy photons emitted by such electrons while they leave the focus.
Such three-fold diagnostics realized in the frame of one experiment could essentially improve the accuracy of measurements, both for the intensity peak amplitude and also for the field distribution in the focal area. 

In this paper, we present results of numerical calculations of electrons and photons energy-momentum distributions for laser parameters corresponding to the current status of the multi-PW Apollon laser facility \cite{apollon_25}.
We change the hitherto unknown parameters of the beam including its degree of focusing and the peak value of the field amplitude in the focus center to trace the evolution of electron and photon distributions triggered by this change. 
To this end, we developed a Monte Carlo (MC) code, which probabilistically simulates tunnel ionization and allows calculating electron trajectories from the instant of birth near the parent ion to the escape from the laser focus.
These trajectories are then used to find the spectra of emitted photons according to the method applied in our earlier work \cite{makarenko_pra26}.
The MC code developed in this work allows greatly accelerating the calculations, compared to a particle-in-cell (PIC) code, because the statement of the problem excludes the necessity of calculation of the back reaction of charges on the electromagnetic field configuration.

The paper is organized as follows.
In the next Section, we describe the statement of the problem and introduce basic equations used for the calculation of electron and x-ray spectra.
Section III describes the numerical set-up including the Monte Carlo code.
Electron and photon spectra and their evolution with the beam peak power and the focal field configuration are discussed in Section IV.
The last Section contains conclusions and outlook.

\section{Statement of the problem}

We consider the dynamics of photoelectrons created through tunnel ionization of argon atoms in the focus of an intense laser pulse and calculate momentum distributions of such electrons after escape from the laser field and angular-resolved spectra of X-ray photons emitted by these electrons during their motion.
Parametrically, we focus on the characteristics of the Apollon multi-PW laser in its present status \cite{apollon_25}:
the FWHM pulse duration is $\tau=22$fs; the pulse energy is $W=70$J; the peak power $P_{\rm max}=3.5$PW at the wavelength $\lambda=0.8\mu{\rm m}$; the expected focal waist is $w_0=2.1\mu{\rm m}$ (considering Gaussian beam; FWHM waist 2.5 mum); the ellipticity is $\rho\approx 0.15$.
At these parameters, the peak intensity in the central part of the focus is ${\cal I}_{\rm max}\approx 3.5\cdot 10^{22}{\rm W/cm}^2$.
However, neither the peak power at the target, nor the waist $w_0$ are accurately known, and the uncertainties have to be minimized using diagnostics, whose possible realization we discuss here.
Thus our aim is to study the sensitivity of the photoelectron and photon distributions to the peak intensity, which we will vary by changing either the peak power $P_{\rm max}$ or the spot size $w_0$.

For the gas target, we choose argon as an element widely used in strong-field ionization experiments. 
Considering the interval of intensities ${\cal I}_{\rm max}=10^{22}-10^{23}{\rm W/cm}^2$, which is likely to be explored at Apollon for the current level of the peak power, xenon and krypton could look more suitable targets \cite{poprz_pra19}, because there are levels there to be efficiently ionized in this intensity interval.
However, for the first proof-of-principle experiments, we suggest argon, which is cheaper, and its ionization potentials allow probing the domain ${\cal I}_{\rm max}=10^{21}-5\cdot 10^{22}{\rm W/cm}^2$, which can be 
reliably achieved with a few-PW level source.
The concentration of atoms is assumed to be low, $n_{at} \simeq 10^{11}$ cm$^{-3}$, which 
allows one to neglect space charge effects and interaction between photoelectrons and other ions.

We consider three configurations of the field.
The first one is a linearly polarized electromagnetic field of a Gaussian laser pulse propagating in the direction of the $x$ axis with zero longitudinal fields.
\begin{equation}
\begin{split}
    {\bf E}({\bf r},t)&=  \frac{{\bf E}_0}{\sqrt{1 + x^2/x_R^2}} 
    \exp\left(-\frac{r^2}{w_0^2\left[1 + x^2/x_R^2\right]}\right) \\
    &\times \cos\left(kx - \omega_L t - \arctan\left(\frac{x}{x_R}\right) + \frac{kr^2}{2x\left[1 + x_R^2/x^2\right]} 
    \right) \\
    &\times f(\omega_L t-kx)~,~~~r=\sqrt{y^2+z^2}~,\\
    {\bf H} &= {\bf n}\times{\bf E}~,~~~{\bf n}=\frac{\bf k}{k}~, 
\end{split}
\label{gauss-beam}
\end{equation}
where $E_0$ is the electric field peak amplitude, $f$ is the time envelope of total duration $\tau=2M\pi/\omega_L\equiv MT_L$, $x_R =\pi w_0^2/\lambda$ is he focal Rayleigh length, $\omega_L$ and $T_L$ are the frequency and the laser period, correspondingly.
The vector ${\bf E}_0$ is directed along the $y$-axis.
The time envelope has the form:

\begin{equation}
f(\omega_L t-kx) =
\begin{cases}
    \cos^2\!\left(\frac{(\omega_L t-kx)}{2M}\right), & \omega_L t-kx \in [-\pi M, \pi M],\\
    0, & \omega_L t-kx \notin [-\pi M,\pi M]~.
\end{cases}
\label{gaussian}
\end{equation}
Note that Eq.(\ref{gauss-beam}) is valid in the paraxial approximation when the pulse duration is sufficiently long, $\omega_L\tau\gg kw_0$ \cite{narozhny_jetp00}.
For the parameters introduced above, one obtains $\omega_L\tau\approx 40$, and $kw_0\approx 20$ so that Eq.(\ref{gauss-beam}) is on the edge of its applicability.

The second field configuration is also a linearly polarized Gaussian pulse but with but with the magnetic field defined as
\begin{equation}
    {\bf H} = -c \cdot \operatorname{rot} \int^t {\bf E}(t^{\prime}) \, dt^{\prime}~,
\label{gauss-long}
\end{equation}
and has a non-vanishing longitudinal component. For the considered parameters, $|H_x|\approx 0.05E_0$.

Finally, the third field configuration is a Gaussian pulse with elliptical polarization at small ellipticity $\rho=0.15$, which we assume relevant for realistic focusing optical elements. 
In this case, the field component $E_y$ is given by Eq.(\ref{gauss-beam}) divided by $\sqrt{1+\rho^2}$, and $E_z({\bf r},t)=\rho E_y({\bf r},t-\pi/2\omega)$. The magnetic field is elliptically polarized accordingly, however, we omit the longitudinal component. This way we study the impact of the longitudial field and ellipticity independently.

The real structure of the electromagnetic field in the focal area is by far more complicated than that of the Gaussian beam. 
A realistic field distribution, which can be obtained by solving numerically the Maxwell equations, e.g. using a PIC code in the particle-free mode \cite{smilei}, demonstrates a diffraction structure in the form of Airy rings.
However, as we are solely interested here in ionization of inner atomic levels with binding energies of hundreds and thousands eV (the ionization potential of the hydrogen-like ${\rm Ar}^{17+}$ is $I_p=4426$eV) and in the dynamics and radiation of photoelectrons accelerated up to relativistic gamma-factors $\gamma\simeq 10^2 - 10^3$, the central part of the focal area, where the field is relativistically strong, is only of interest for our calculations.
Significance of relativistic effects is essentially determined by the value of the dimensionless invariant field amplitude 
\begin{equation}
    \label{a0}
    a_0=\frac{eE_0}{mc\omega_L}~,
\end{equation}
which is $a_0=10^2$ for a linearly polarized field of wavelength $\lambda=1\mu{\rm m}$ and intensity ${\cal I}=2\cdot 10^{22}{\rm W/cm}^2$.
For the description of the central part of the focal area, the approximation of Gaussian beams works reasonably well for $w_0/\lambda\approx 2-3$ and higher.
For tighter focusing, different exact solutions for stationary beams \cite{narozhny_jetp00,skupin_jcp16} can be used instead of the Gaussian family.

\subsection{Tunnel ionization}
Here we introduce the analytic formalism for description of the ionization step.
The model is based on the theory of non-relativistic ionization in static and low-varying electromagnetic fields proposed by Smirnov and Chibisov (SC) \cite{smirnov_jetp66} and Perelomov, Popov and Terentiev (PPT) \cite{popov_jetp66,perelomov_jetp67,popov_usp04,poprz_jpb14}.
In this subsection, we use atomic units $e = m =\hbar = 1$.

Ionization proceeds in the tunnel regime specified by smallness of the Keldysh parameter \cite{keldysh_jetp1965, popov_usp04}.
\begin{equation}
    \gamma_{\rm K}=\frac{\sqrt{2I_p}\omega_L}{E_0}~,
    \label{gamma-K}
\end{equation}
For laser fields with wavelengths $\lambda \simeq 1 \ \mu$m and intensities $I \gtrsim 10^{18}$ W/cm$^2$ $\gamma_K \simeq 10^{-2}-10^{-3}$  (see \cite{poprz_pra19} for estimates), so that the tunnel regime is well secured. 
Relativistic effects in the rate of ionization play no significant role up to intensities $\simeq 10^{24}$ W/cm$^2$ \cite{poprz-jetpl23,baranov_pra26}. 

Under these conditions, the ionization probability is determined by the instant electric field amplitude at the $j$-th atom position $\boldsymbol{E}(\boldsymbol{R}_j,t)$. 
Atoms are considered at rest during the femtosecond pulse duration.
We assume that the electrons are removed sequentially with increasing ionization potentials. 
Deviations from this simplest sequence of ionization were identified in a recent paper \cite{mironov_pre25}.
However, they do not considerably affect the number of photoelectrons released at a given intensity.
According to the SC and PPT theories \cite{smirnov_jetp66,perelomov_jetp1967,popov_usp04}, the probability for the outer electron with orbital and magnetic numbers $l$ and $m$ to be removed from its atom (ion) with residual charge { $Z_c$ (in these notations, $Z_c=1$ for a neutral atom and $Z_c=N$ for $A^{(N-1)+}$ ion)} during the time interval $dt$ is given by the  PPT static rate \cite{smirnov_jetp66,perelomov_jetp1966,perelomov_jetp67,popov_usp04}
\begin{equation}
\begin{gathered}
    dW(t) = w(I_p,\nu,l,m; E(t))dt~, \\
    w = 2^{2\nu+1}C^2_{\nu l} B_{lm} I_p F^{1+|m|-2\nu}(t) \exp\left( -\frac{2}{3F(t)} \right)~.
\end{gathered}
\label{PPT}
\end{equation}
Here
\begin{equation}
  F(t) = \frac{\sqrt{{\bf E}^2(t)}}{(\rm 2 I_p)^{\rm 3/2}}  
  \label{F}
\end{equation}
is the reduced electric field scaled by the characteristic field $E_{\rm ch}=(2I_p)^{3/2}$ of the atomic level,
\begin{equation}
    \nu = \frac{Z_c}{\sqrt{2I_p}}
    \label{nu}
\end{equation}
is the effective principal quantum number,
\begin{equation}
    C^2_{\nu l} = \frac{2^{2\nu-2}}{\nu\Gamma(\nu+l+1)\Gamma(\nu-l)}
    \label{C}
\end{equation}
is the squared asymptotic coefficient of the bound state single-electron wave function calculated in Hartree approximation \cite{popov_usp04} and the factor
\begin{equation}
    B_{lm} = \frac{(2l+1)(l+|m|)!}{2^{2|m|}|m|!(l-|m|)!}
    \label{B}
\end{equation}
reflects the angular structure of the bound state.

For the simulations of this work, we secure several important limitations of the SC and PPT rates.
Firstly, we focus on dynamics and radiation of high-energy photoelectrons produced in the central part of the focus, where the maximal local laser intensity is close to its peak value.
This allows us to consider only ionization of the inner shells of argon, starting from the 2p levels.
By this, we exclude from consideration the low-energy electrons, which predominantly appear due to ionization of the outer shell in a large low-intensity part of the focal volume.
Secondly, the 22fs pulse duration is still sufficient to make sequential tunneling the main ionization mechanism at high intensities \cite{tyurin_arxiv25}.
Finally, by modeling the ionization step (see Section III below for details) we check that the given level is being depleted before the reduced field (\ref{F}) reaches 0.1.
For higher values of $F$ the rate (\ref{PPT}) highly overestimates the probability of ionization, and empirical extensions of (\ref{PPT}) have to be used to secure quantitative accuracy \cite{popov_usp04,tong_jphysb2005,kostyukov_pra2018}.

\subsection{Post-ionization photoelectron dynamics and radiation}

The post-ionization motion of photoelectrons is described by solving numerically  the classical equation of motion
\begin{equation}
    \frac{d{\bf p}}{dt}=e\bf{E} + \it\frac{e}{c} \bf{v}\times{\bf H}
    \label{eqm}
\end{equation}
with the electric and magnetic fields given by Eqs. \eqref{gauss-beam}--(\ref{gauss-long}). 
Initial conditions are determined by the tunnel ionization mechanism and take into account that the tunnel barrier width is small compared to the laser wavelength:  
\begin{equation}
    \label{ic}
    {\bf r}(t_{jn})={\bf R}_j, ~~~{\bf v}(t_{jn})=0.
\end{equation} 
Here, $j$ denotes the atom number and $n$ that of the electron in this atom. 

Photoelectron trajectories corresponding to the initial conditions (\ref{ic}) in extremely strong fields are of specific spatial shape \cite{kalymbetov_qe21,makarenko_pra26}.
Due to a strong acceleration along the propagation direction of the laser pulse, the electrons make a few or even no single full oscillation before leaving the laser focus.
For lower ionization potentials, the electrons predominantly leave the focus through its lateral surface, while the electrons removed  from the inner shells at time instants when the laser field was close to its maximum, experience no oscillations and leave the focus through its back surface.

In a plane wave of finite duration, the initial condition (\ref{ic}) leads to a simple relation between the asymptotic value of the longitudinal $p_x$ and lateral ${\bf p}_{\bot}$ photoelectron momenta \cite{kalymbetov_qe21,makarenko_pra26}. In terms of the polar emission angle $\theta={\rm tan}^{-1}(p_{\bot}/p_x)$ and the relativistic gamma-factor $\gamma$ this relation reads
\begin{equation}
    \label{gamma-theta}
    \gamma \approx\frac{2}{\tan^2\theta}.
\end{equation}
Note that in a linearly polarized field of type (\ref{gauss-beam}) the photoelectron motion is confined in the $(x,y)$ plane, despite the focusing.
In fields with a longitudinal component (\ref{gauss-long}) or with nonzero ellipticity, the electron trajectories become three-dimensional, which may reduce the accuracy of estimate (\ref{gamma-theta}), see Section IV for results and discussion.

The spectral-angular distribution of radiation emitted by an electron moving along the trajectory ${\bf{r}}_{jn}(t)$ is calculated using the pivotal expression of classical electrodynamics \cite{Jackson,Landau2}
\begin{equation}
    \frac{d{\cal{E}}_{jn}}{d\omega dO_{\bf n}}=   \frac{e^2\omega^2}{4\pi^2c^3} \Bigg|\int\limits_{t_{jn}}^{\infty} dt~ [{\bf n} \times [{\bf n} \times {\bf v}_{jn}(t) ]]e^{i\omega[t-{\bf n}\cdot{\bf r}_{jn}(t)/c]} \Bigg|^{\rm 2}~.
    \label{sad}
\end{equation}
Here $dO_{\bf n}$ is the solid angle in the direction defined by the unit vector ${\bf n}$.

\section{Numerical set-up and method}

The numerical method consists of two sequential stages: (a) Monte-Carlo modeling of ionization and (b) numerical integration of electron trajectories and calculation of the radiation integrals along these trajectories. 
Atoms are randomly distributed in a rectangular box centered at the focal spot. The box size is $110w_0$ along $x$ axis and $16w_0$ in the lateral direction, with $w_0 = 2.65\lambda$. 
Such an extended computational volume allows one to ensure that all parts of the space where the $2p$ and lower shells can be ionized at the peak intensity $\simeq10^{23} \, \mathrm{W}/\mathrm{cm}^2$ in the center are included. 
Assuming the concentration $n_{at} \simeq 10^{11} \, \mathrm{cm}^{-3}$, we set the total number of atoms in the box equal to $N_a=5\cdot 10^4$. 
For the initial configuration, we chose the ground $2p^62s^21s^2$ state of $\mathrm{Ar}^{8+}$.
This helps to exclude ionization of the $3p^63s^2$ shells which proceeds at low intensities beyond the tunneling regime and gives rise to low energy photoelectrons, which are out of interest for this study.

We consider multiple ionization as a sequence of independent steps, which allows us to use parallel methods and speeds up the numerical calculation. 
The electric field at the position of the atom is evaluated at discrete time steps $\omega_L\Delta t = 0.001$, and the ionization probability of the considered level is calculated using the rate (\ref{PPT}) multiplied by this time interval securing the condition that this probability always appears small compared to unity. 
Then a random number uniformly distributed in the interval $[0,1)$ is generated, and an ionization event is registered when this number is smaller than the calculated probability. 
After a successful ionization event, we update the ionic state, and the atom position ${\bf R}_j$ and the instant of time $t_{jn}$ become the initial conditions for the post-ionization electron motion. 

To find the electron trajectory ${\bf r}_{jn}(t;{\bf R}_j,t_{jn})$ we solve the equation of motion (\ref{eqm}) rewritten in the form of two first-order equations by Runge-Kutta (RK) method of order 4(5). 
It is realized by Python solve\_ivp method, where Runge-Kutta is one of available options. For parallel calculations, the pool.imap is utilized. 

First, we calculate the final momenta of the electrons by integrating their equations of motion from the instant of ionization to $\omega_L t_{\rm out} = 10^4$, without storing information about the trajectories. 
The laser pulse leaves the target at $\omega_Lt = 10^3$, so the integration continues long after that. 
The trajectories are recorded at uniformly spaced time points with a step $\omega_L \delta t = 0.1$. 
Under RK4(5) method the error is controlled assuming accuracy of the fourth-order method, but steps are taken using the fifth-order accurate formula, so the internal integration step size is independent of the time interval $\delta t$ used to record the trajectory data. We use the final momenta arrays for plotting angle-resolved photoelectron spectra and the trajectories for the calculation of the emission spectra.

\section{Results and discussion}

\subsection{Photoelectron spectra}

We start from a discussion of the photoelectron distributions in the $(\gamma,\theta)$ plane with $\theta$ being the angle between the propagation direction of the beam and the photoelectron final momentum.
In a linearly polarized plane wave pulse of finite duration and for $a_0\gg 1$, the $\gamma$ factor of an escaping electron is connected to the angle according to (\ref{gamma-theta}).
This function itself does not depend on $a_0$ and therefore bears no information on the peak laser intensity.
Instead, the minimal angle $\theta_{\rm min}$ clearly seen for the distributions shown in the upper and lower rows of Fig.1, depends on the value of $a_0$ 
Within this plane wave model, for an electron born at the peak value of the field,
\begin{equation}
    \label{gamma-max}
    \theta_{\rm min}\approx\frac{2}{a_0},~~~\gamma_{\rm max}\approx\frac{a_0^2}{2}
\end{equation}
Blue solid lines in Fig.1 show the analytic scaling (\ref{gamma-theta}), which, for the fields without a longitudinal component, reproduce the high energy parts of the distributions remarkably well. 

In all cases, there are a low-energy $(\gamma<2)$ and a mid-energy $(\gamma<10^2)$ haloes populated by the electrons escaping through the lateral surface of the focus \cite{makarenko_pra26}.
Note that the focal parameter $w_0=2.7\lambda\approx 2.2\mu{\rm m}$ is quite small, so that the approximation of such a focus by a plane wave is essentially rough.
However, results in Fig.1 (a-c) and (g-i) show that for entirely transverse field configurations this approximation describes the electron dynamics quite accurately, for the high-energy part of the spectrum.
Thus, observation of high-energy electrons near the bounds (\ref{gamma-max}) could provide a better diagnostics of the field near the focal center.

In contrast, the field distribution with a longitudinal magnetic component (\ref{gauss-long}) accelerates the electrons in a different way, so that the angular spread of high-energy electrons appears much broader, and the scaling (\ref{gamma-theta}) no longer gives a qualitatively correct description of this part of the spectrum.
An experimental observation of such angular-resolved spectra in their high-energy $\gamma >10^2$ part should in particular give a clear evidence of the presence of longitudinal fields and their relative magnitude.

\begin{figure}
    \centering
    \includegraphics[width=\linewidth]{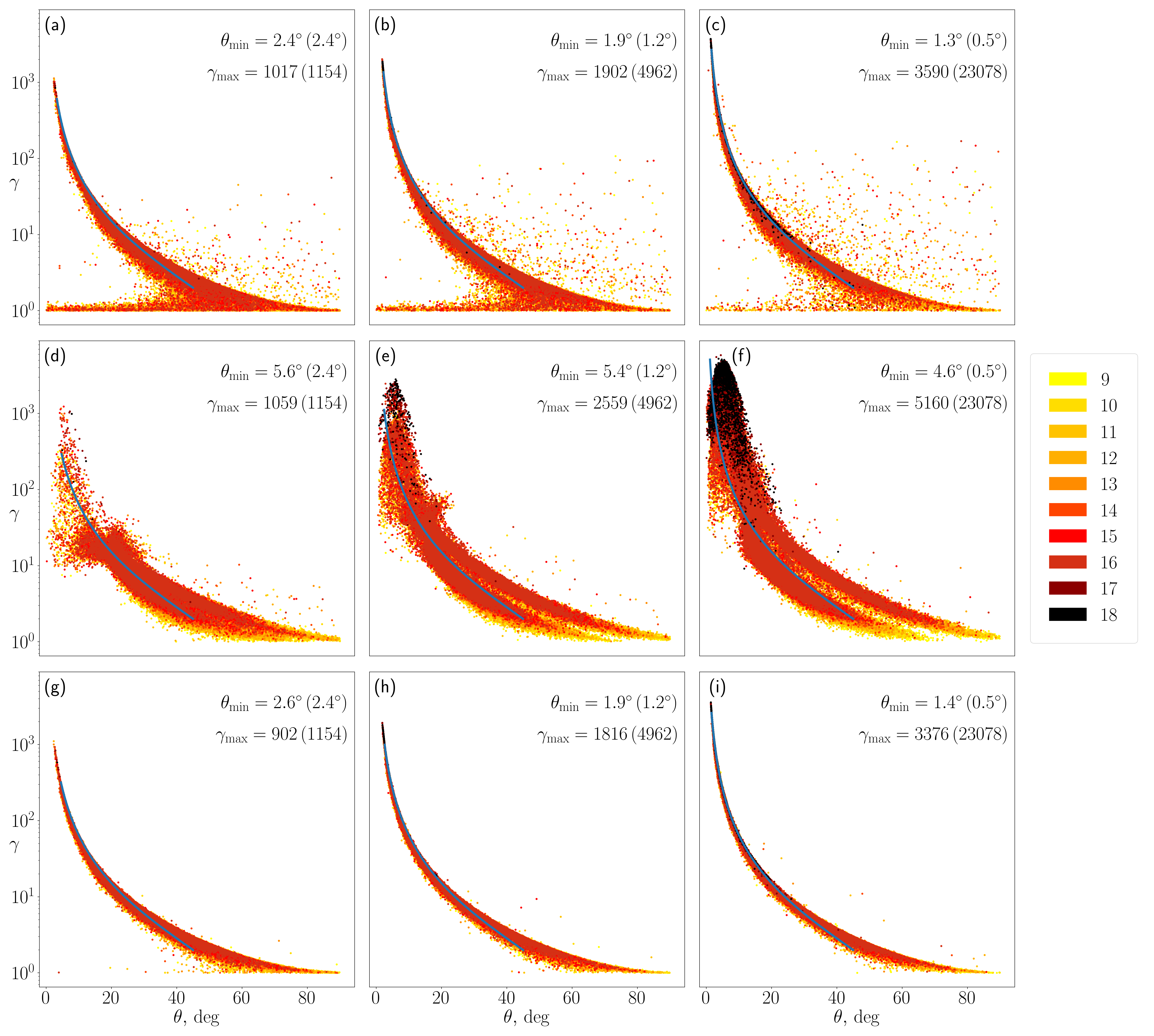}
    \caption{Photoelectron $(\gamma,\theta)$ distributions for different peak intensities and field configurations. The peak intensity is ${\cal I}=5 \times 10^{21} \, \mathrm{W/cm^2}$ (left column), ${\cal I}=2.15 \times 10^{22} \, \mathrm{W/cm^2}$ (middle column) and ${\cal I}=1 \times 10^{23} \, \mathrm{W/cm^2}$ (right column). The field configuration is: linear polarization, no longitudinal field (\ref{gauss-beam}) (upper row), linear polarization with a longitudinal field (\ref{gauss-long})(middle row), elliptical polarization with $\rho = 0.15$ and no longitudinal field (lower row). The plane wave dependence (\ref{gamma-theta}) is shown by a solid blue line. The number of the argon shell is color-coded starting from $2p^6$ (light-yellow) to $3s^1$ (black). Note that a considerable contribution of $3s$ electrons is only present at peak intensities ${\cal I}>2\cdot 10^{22}{\rm W/cm}^2$ (middle and right columns).
    In each panel, the values of $\gamma_{\rm max}$ and $\theta_{\rm min}$ are shown as extracted from the simulation results and (on parentheses) calculated along (\ref{gamma-max}).}
    
    \label{fig:electron_spectra}
\end{figure}
The main message of the spectra shown in Fig.1 and Fig.2 is in the sensitivity of $\gamma_{\rm max}$ and $\theta_{\rm min}$ to the peak laser intensity and in the dependence of this sensitivity on the field configuration in the focus.
As is seen from the numbers shown on the panels, a good agreement between the theory and simulations takes place at the lowest intensity.
This can be explained by the fact that, in contrast to (\ref{gamma-theta}), which does not contain $a_0$, the limiting values (\ref{gamma-max}) do, so that the theoretical estimates are valid under assumption that the most energetic electrons were created at the peak intensity. 
This is the case for ${\cal I}< 5\cdot 10^{21}{\rm W/cm}^2$, the value determined by the highest ionization potential of argon \cite{poprz_pra19}. 
A position of the spectral cut-off $\gamma_{\rm max}$ appears the most reliable measure of the peak intensity even in the presence of the longitudinal field (see (a), (d) and (g) in Fig.1), but under a condition that this intensity is close to the ionization threshold for highly charged ions of argon.

\begin{figure}
    \centering
    \includegraphics[width=\linewidth]{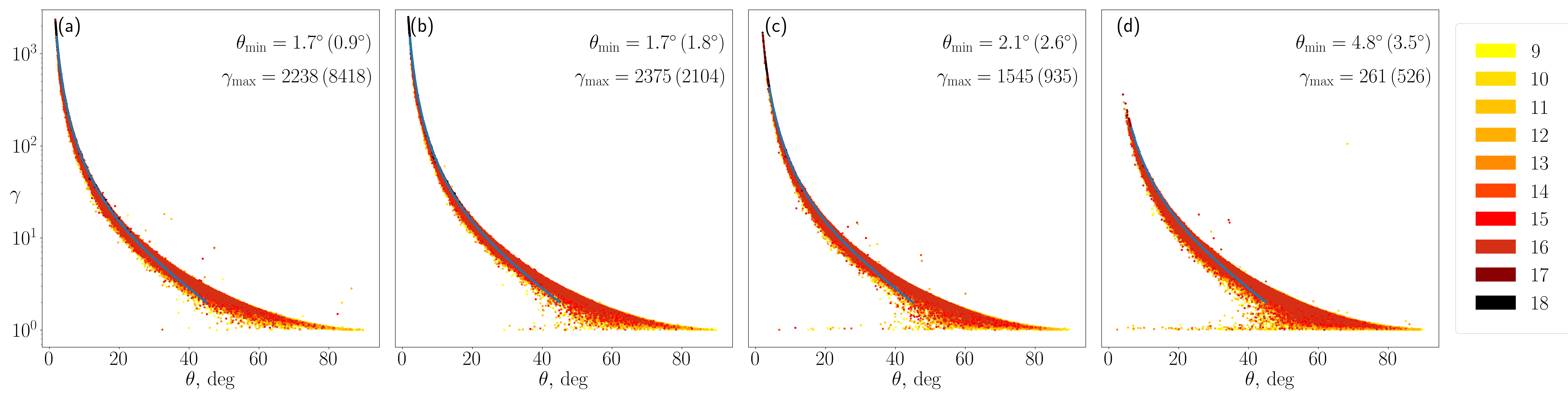}
    \caption{Same as in Fig.1(h) for  different values of the beam waist, $w_0=2.7\lambda$ (a), $2w_0$ (b), $3w_0$ (c) and $4w_0$ (d) at fixed pulse energy $W=70 \ \mathrm{J}$ and duration $\tau=22$ fs. The smallest waist (a) corresponds to the peak intensity ${\cal I}=2.15\cdot 10^{22} \ {\rm W/cm}^2$, the largest one (d) -- to ${\cal I}=2.28\cdot 10^{21} \ {\rm W/cm}^2$.}  
    \label{fig:different w0 electron spectra}
\end{figure}

A complementary diagnostics with photoelectrons can be developed by measuring their angular distributions in the plane perpendicular to the beam propagation direction.
Such distributions integrated over the energy interval $\gamma=10^2..10^3$ are shown in Fig.3 with angle $\theta$ as a radial coordinate and the azimuthal angle $\varphi$ -- as an azimuthal coordinate. 
Intensities and field configurations correspond to those of the second and the third rows of Fig.1.
In the case of a linearly polarized transversal field, the electron motion is precisely confined in the $({\bf E}_y,{\bf k})$ plane so that the angular distributions appear one-dimensional.
The central message of this plot is in a strong influence of the longitudinal field on the distributions.
Even being as small as 0.05 of the transversal field amplitude, it leads to a considerable spread of the distributions, particularly, at the highest intensity.
The distributions shown in the third row and corresponding to the field parameters expected at the Apollon laser facility \cite{apollon_25} demonstrate a clear spread along the angle $\varphi$ with the peak intensity increasing.

\begin{figure}
\includegraphics[width=\linewidth]{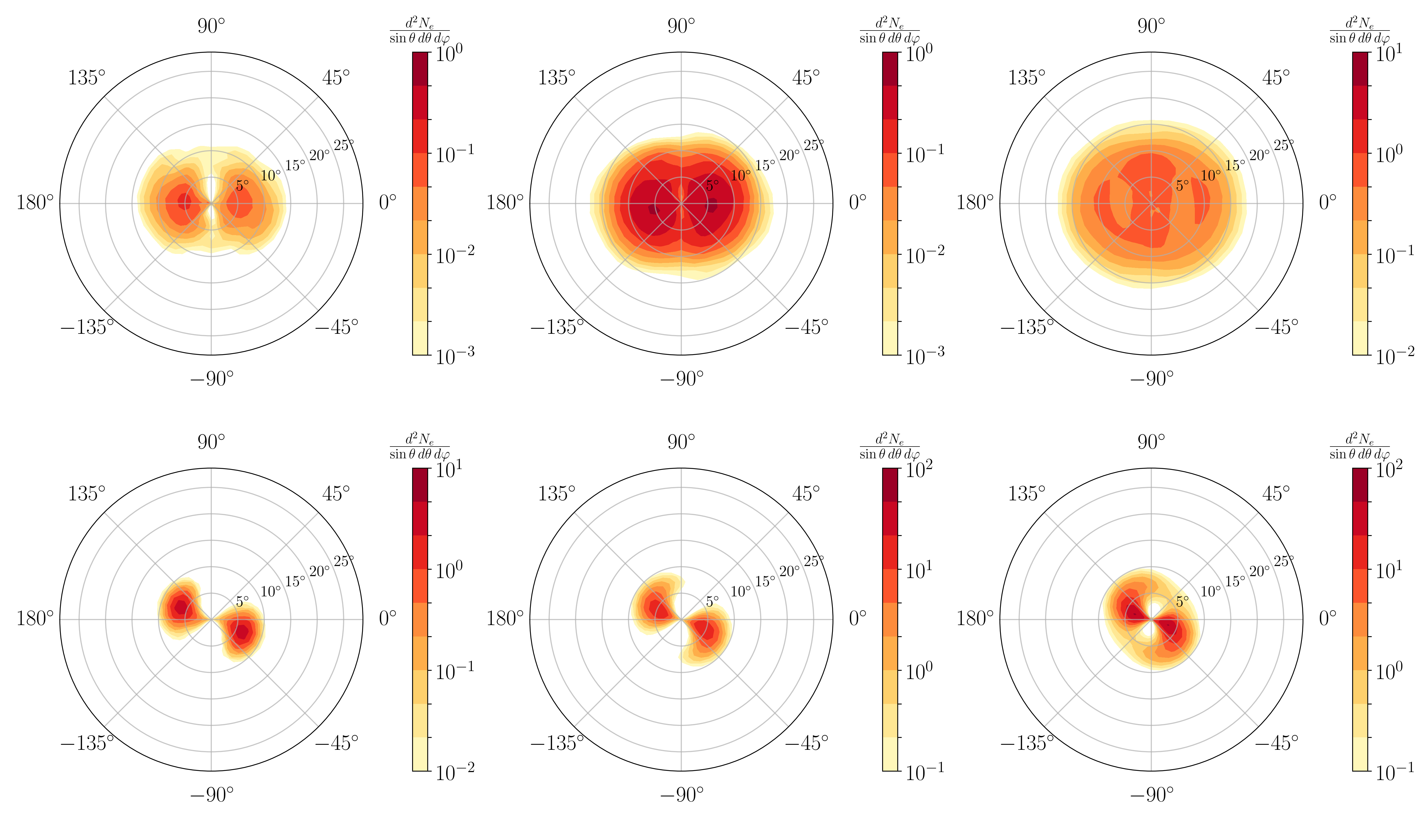}
\caption{Angular distributions of photoelectrons integrated over energy in the range $\gamma= 10^2..10^3$. The angle $\theta$ between the propagation direction of the beam and the electron momentum is shown as a radial coordinate; the azimuthal angle $\varphi$ -- as an angular coordinate. The panels correspond to those of the second and the third rows of Fig.1.}
\label{fig:Angle_distr_e}
\end{figure}

\subsection{Radiation spectra}

Finally, using Eq.(\ref{sad}) we calculate angular distributions of the emitted radiation.
The distributions are integrated in the frequency interval $10^2\omega_L<\omega<3\cdot 10^2\omega_L$, where the maxima of the synchrotron-like spectra of radiation are located \cite{makarenko_pra26} for all intensities selected for our calculations.
As in the plots discussed above, it is clearly seen that the longitudinal field (second row) gives rise to diffuse distributions with a broader spread over the azimuthal angle $\varphi$ and a weaker dependence on the peak laser intensity.
Apart from this feature, the distributions of the upper (linear polarization) and the lower (elliptical polarization) rows demonstrate a clear tendency of narrowing with the intensity increasing.
Note that the distributions of Fig.4 are normalized, so that a growth in the integral photon signal with the intensity increasing is not reflected.

The total number of photons emitted by the whole ensemble of atoms in the focal volume in the selected frequency interval can be estimated as \cite{makarenko_pra26} $N_{\gamma}\simeq 10^2-10^4$.
Although the brightness is not high, such photon flux can be experimentally detected and resolved in angles\footnote{
Note that our previous work \cite{makarenko_pra26} contains an erroneous statement that the total number of photons is 2-3 times higher than that given by Eq.(27) there. 
The correct formulation is that the total number of high-energy photons emitted in the whole spectral range per laser pulse is $2\times10^3 - 3\times 10^4$, which corresponds with the results of current work.}.

The photon diagnostics provides a probe, which is  complementary to that based on the measurement of photoelectrons.
In particular, both the photoelectron and the photon angular distributions appear highly sensitive to the presence of the longitudinal field component.
For purely transverse fields, the width of the photon angular distribution can be used as measure of the peak intensity in line with the values $\gamma_{\rm max}$ and $\theta_{\rm min}$ extracted from angular-resolved spectra of photoelectrons.

\begin{figure}
\centering
\includegraphics[scale = 0.5]{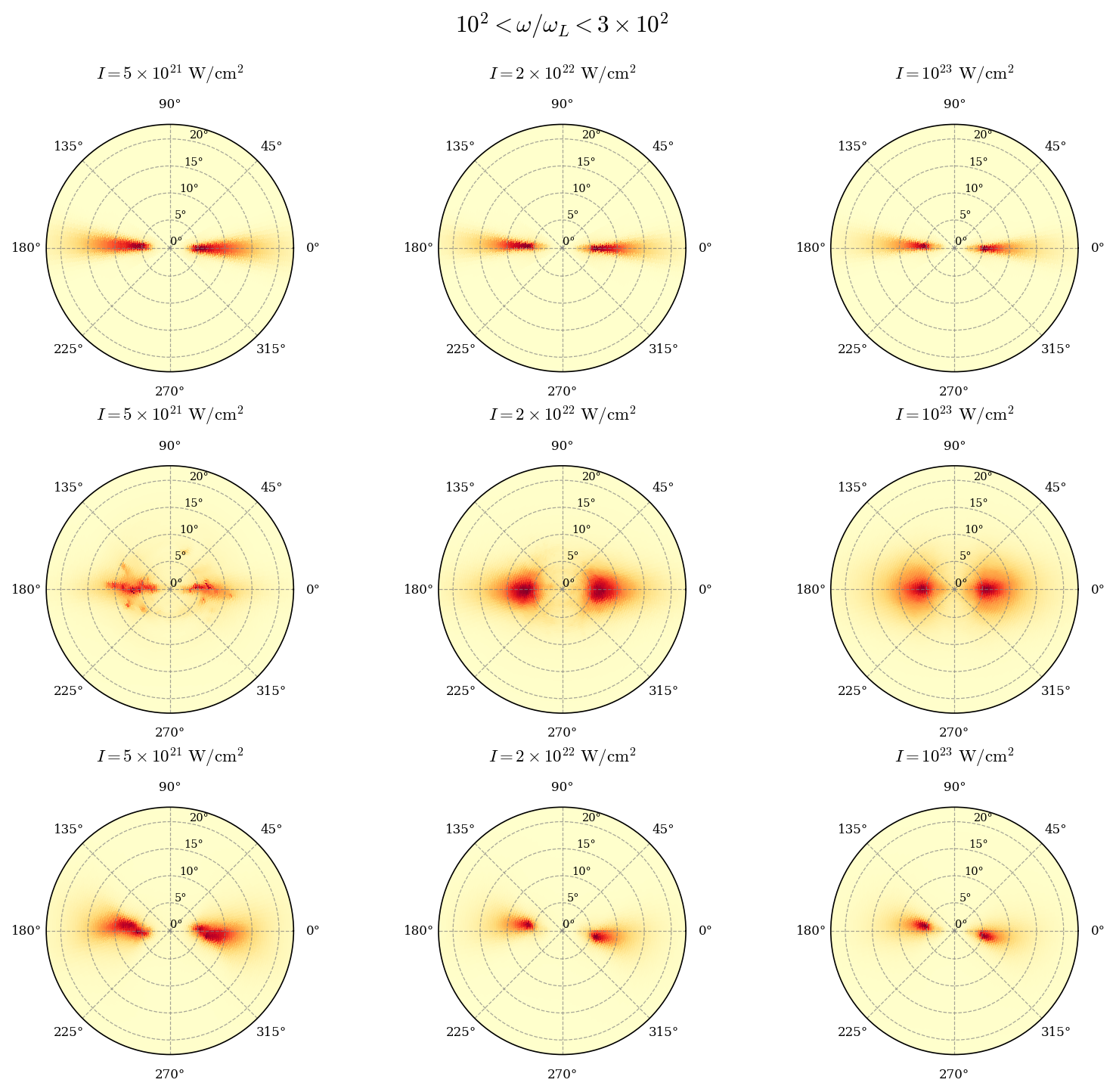}
\caption{ Angular distributions of the emitted photons in the frequency interval $10^2 < \omega/\omega_L < 3 \times 10^2$ for laser intensities $\mathcal{I} = 5 \times10^{21}$ W/cm$^2$ (left column), $\mathcal{I} = 2 \times10^{22}$ W/cm$^2$ (middle column), $\mathcal{I} = 1 \times10^{23}$ W/cm$^2$ (right column) and different field configurations: Gaussian beam (upper row), Gaussian beam with longitudinal field (middle row), elliptically polarized Gaussian beam (lower row). Azimuthal angle $\varphi$ is measured in the plane perpendicular to the laser wave vector $\mathbf{k}$, relative to the electric field vector $\mathbf{E}$ directed along the $y$-axis in the case of linear polarization; the radial distance represents the angle $\theta$ between $\mathbf{k}$ and the photon propagation direction. All distributions are normalized to unity for better visibility.}
\label{fig:Angle_distr}
\end{figure}

\section{Conclusions and outlook}

Summarizing, we have presented results of numerical calculations of photoelectron and X-ray photons energy-angular distributions for the case, when the photoelectrons are produced in the central part of a focused high-power laser beam with parameters close to the currently achieved at several multi-PW laser facilities, including Apollon \cite{apollon_25}, whose beam energy and pulse duration have been chosen as the core parameters of the calculation.
The distributions are shown to be essentially sensitive both to the peak laser intensity and to the field configuration in the focus.

Our results show that the presence of longitudinal field components should be considered as the main obstacle for the reliable intensity diagnostics.
At the same time, both the photon and the electron angular distributions are shown to be efficient tools for the estimation of the longitudinal field relative amplitude.

The values of $\gamma_{\rm max}$ and $\theta_{\rm min}$ appear good measures of the peak intensity, when there are levels with ionization potentials, which correspond efficient ionization at such intensities.
This statement holds, although with lower quantitative accuracy, even in the presence of longitudinal fields
This returns us to the problem of the intensity-dependent composition of the gas target for ionization-based diagnostics of extreme laser fields \cite{poprz_pra19,peganov_mre20}.

A correlative analysis of momentum and angular distributions of photons and photoelectrons by means of neural networks may be helpful to make considerable further advances of this problem.

\section*{Acknowledgment}
AAM and SVP acknowledge fruitful discussions with R. Ta\"{i}eb and S. Turban.
NVM and SVP acknowledge support of the Russian Science Foundation through Grant No.25-22-00308.

\vskip 1cm
\providecommand{\newblock}{}

 \end{document}